\documentclass[conference, hidelinks]{IEEEtran}
\IEEEoverridecommandlockouts
\usepackage{amsmath,amssymb,amsfonts}
\usepackage{algorithmic}
\usepackage{graphicx}
\usepackage{textcomp}
\usepackage{tikz-uml}
\usepackage{pgf-umlsd}
\usepackage{xcolor}
\usepackage{orcidlink}
\usepackage{svg}
\usepackage[caption=false]{subfig}
\usepackage{paralist}

\def\BibTeX{{\rm B\kern-.05em{\sc i\kern-.025em b}\kern-.08em
    T\kern-.1667em\lower.7ex\hbox{E}\kern-.125emX}}

\newcommand{\upp}{\vspace*{-0.5em}}
\usepackage{url}
\usepackage[%
  backend=biber,
  url=false,
  doi=false,
  isbn=false,
  style=ieee,
  maxnames=3,
  minnames=2,
  maxbibnames=3,
  giveninits,
  sorting=none,
  citestyle=numeric-comp,
  uniquename=init]{biblatex} %

\newif\ifdraft
\ifdraft
\newcommand{\note}[1]{ {\textcolor{blue} { **: #1 }}}
\newcommand{\alnote}[1]{ {\textcolor{red} { ***Andre: #1 }}}
\newcommand{\prnote}[1]{ {\textcolor{teal} { ***Philipp: #1 }}}
\newcommand{\jknote}[1]{ {\textcolor{blue} { ***Johannes: #1 }}}
\newcommand{\marwanote}[1]{ {\textcolor{yellow} { ***Marwa: #1 }}}
\newcommand{\flonote}[1]{ {\textcolor{orange} { ***Florian: #1 }}}
\newcommand{\crnote}[1]{ {\textcolor{purple} { ***Carlos: #1 }}}
\newcommand{\rudinote}[1]{ {\textcolor{green} { ***Rudi: #1 }}}
\newcommand{\ieeenote}[2]{ {\textcolor{blue} { ***IEEE Reviewer #1: #2 }}}
\else
\newcommand{\note}[1]{}
\newcommand{\alnote}[1]{}
\newcommand{\prnote}[1]{}
\newcommand{\jknote}[1]{}
\newcommand{\crnote}[1]{}
\newcommand{\flonote}[1]{}
\newcommand{\marwanote}[1]{}
\newcommand{\jrfnote}[1]{}
\newcommand{\rudinote}[1]{}
\newcommand{\ieeenote}[2]{}
\fi
 
\begin{document}
\bstctlcite{IEEEexample:BSTcontrol}
\title{Application-Oriented Benchmarking of\\ Quantum Generative Learning Using QUARK}

\author{\IEEEauthorblockN{Florian J. Kiwit\IEEEauthorrefmark{1}\IEEEauthorrefmark{3}\IEEEauthorrefmark{4},
Marwa Marso\IEEEauthorrefmark{1}\IEEEauthorrefmark{2}\IEEEauthorrefmark{4}, 
Philipp Ross\IEEEauthorrefmark{1} \orcidlink{0000-0002-4720-9835}, Carlos A. Riofr\'io\IEEEauthorrefmark{1}\orcidlink{0000-0002-7346-9198}, Johannes Klepsch\IEEEauthorrefmark{1}\orcidlink{0000-0002-8247-9590}, Andre Luckow\IEEEauthorrefmark{1}\IEEEauthorrefmark{3}\orcidlink{0000-0002-1225-4062}}
\IEEEauthorblockA{\IEEEauthorrefmark{1}BMW Group, Munich Germany\\
                  \IEEEauthorrefmark{2}Technical University Munich, Germany\\
                  \IEEEauthorrefmark{3}Ludwig Maximilian University Munich, Germany\\
                  \IEEEauthorrefmark{4}{\footnotesize Authors contributed equally}}}

\maketitle
\begin{abstract}
Benchmarking of quantum machine learning (QML) algorithms is challenging due to the complexity and variability of QML systems, e.\,g., regarding model ansatzes, data sets, training techniques, and hyper-parameters selection. The \emph{QUantum computing Application benchmaRK} (QUARK) framework simplifies and standardizes benchmarking studies for quantum computing applications. Here, we propose several extensions of QUARK to include the ability to evaluate the training and deployment of quantum generative models. We describe the updated software architecture and illustrate its flexibility through several example applications: (1) We trained different quantum generative models using several circuit ansatzes, data sets, and data transformations. (2) We evaluated our models on GPU and real quantum hardware. (3) We assessed the generalization capabilities of our generative models using a broad set of metrics that capture, e.\,g., the novelty and validity of the generated data. 
\end{abstract}

\begin{IEEEkeywords}
quantum computing, benchmark, machine learning, generative modeling
\end{IEEEkeywords}

\section{Introduction} 

\emph{Motivation:}
Quantum computing promises to accelerate certain computational kernels relevant to a broad set of industry applications~\cite{quantum_industry_applications}. Quantum Machine Learning (QML)~\cite{schuld_machine_2021} integrates the discipline of machine learning, i.\,e., the development of algorithms that enable computers to learn from data, and quantum computing that potentially enables more powerful models~\cite{PhysRevX.12.021037, riofrio2023performance}.

There are many relevant industry problems, e.\,g., data augmentation for machine learning systems  (e.\,g., for computer vision classification systems that monitor manufacturing defects~\cite{Guala2023}), synthetic data generation as replacement for sensitive data~\cite{Watkins2023}, for risk and portfolio management~\cite{Woerner2019}, and generative design~\cite{10.1115/1.4053859}. Further, QML-enhanced optimization approaches have broad applications in various domains~\cite{alcazar2022geo}. Generative QML techniques, such as Quantum Circuit Born Machines (QCBMs)~\cite{e20080583, Benedetti_2019} and Quantum GANs (QGANs)~\cite{Dallaire_Demers_2018}, are promising algorithms that demonstrated comparable training performance to their classical counterparts, requiring fewer parameters~\cite{riofrio2023performance}.

\emph{State-of-the-art and limitations:} 
Studying the performance of quantum computers and comparing them to their classical counterparts is an essential part of current research. Benchmarks are standardized tests comprising workloads, datasets, and metrics for measuring the performance of a system. Various benchmarks have been proposed on different levels, e.\,g., quantum volume~\cite{Cross2019}, CLOPS~\cite{wack2021quality}, and gate fidelity on system-level~\cite{Huang_2019}. Higher-level benchmarks include, e.\,g., the Q-Score~\cite{qscore_2023}, SupermarQ~\cite{supermarq} and the QED-C~\cite{Lubinski_etal_2021, Lubinski_etal_2023} benchmarks.

Application benchmarks that evaluate the workload of a specific real-world application are critical to assess the end-to-end performance of emerging QML systems. Performance and resource estimates for real-world workloads are also important to assess the commercial applicability. In the field of QML, references~\cite{Hamilton_2019, Benedetti_2019, west2022benchmarking, PhysRevResearch.3.023010} are examples of application-level benchmarks (see Section~\ref{sec:qml_bench}). However, these benchmarks do not assess the end-to-end application performance, as they only focus on individual quantum subroutines.

Creating representative benchmarks is challenging. Application-level metrics are often difficult to compare~\cite{amico2023defining} and exhibit complex relationships. For example, quantum error mitigation trades improved solution quality with increased computational demands and longer time-to-solutions. We make a case for an application-level benchmarking framework that can holistically capture workloads, data, and metrics in a unifying context allowing the investigation of such trade-offs.

\emph{Previous work:} Fin\v{z}gar et al.~\cite{Finzgar_2022} introduced the \emph{QUantum computing Application benchmaRK} (QUARK) framework, an open-source~\cite{quarkGithub} framework for designing, implementing, executing, and analyzing benchmarks. QUARK addresses critical requirements of application-level benchmarks, such as implementing realistic workloads and datasets at different scales, supporting multiple implementations, and enabling reproducible results. While the initial architecture of QUARK was designed for optimization benchmarks, other application areas, such as QML, have more complex requirements. Specifically, ML workflows comprise complex datasets, pre-/post-processing routines, hybrid algorithms involving quantum and classical tasks, and metrics. Modularity and flexibility are critical requirements as benchmarking problems and metrics in ML evolve rapidly~\cite{stanford_ai_2023}.

\emph{Key insights, contributions, and artifacts:} The training of QML models involves complex trade-offs between, e.\,g., circuit/model ansatzes, datasets and transformations, complex hybrid algorithms with various hyper-parameters, and the need to optimize both quantum and classical resources~\cite{DBLP:journals/corr/abs-2001-08361}. A framework such as QUARK can prove invaluable for the quantum practitioner. In this work, we make three significant contributions: 
\begin{inparaenum}[(1)]

\item \emph{QUARK 2.0:} We introduce an updated QUARK architecture (see Section~\ref{sec:quark2}), which allows greater flexibility for the definition of benchmarking modules and metrics. QUARK addresses different complexities of creating benchmarks, e.\,g., resource heterogeneity, the standardization of measurements and data collection, and reproducibility and supports the orchestration of complex benchmarking pipelines. It enables benchmark developers to extend QUARK to further applications while retaining support for all existing benchmarking applications. 

\item \emph{QML Benchmarks:} We demonstrate the flexibility of the QUARK framework by adding a new application workflow for QML. We implement a quantum generative model~\cite{riofrio2023performance}, including different modules for data transformations, quantum circuit architectures, and training methods (see Section~\ref{sec:qml}).

\item \emph{Experimentation on GPU and Quantum Hardware:} We present results for the QML benchmark workflow on real hardware and simulations (see Section~\ref{sec:experiments}). We study the training and inference performance of QML models on state-of-the-art high-performance and quantum computing infrastructures (see Section~\ref{sec:training_performance}). Further, we deploy our QML benchmark workflow on real quantum hardware, the IonQ Harmony machine, evaluating the impact of noise (see Section~\ref{sec:hardware}).

\end{inparaenum}

\emph{Limitations:} %
The QUARK benchmarking framework currently focuses on applications in optimization and QML. While the new architecture facilitates the addition of specialized benchmark applications, e.\,g., for quantum-enhanced numerical simulations or quantum chemistry, no blueprint or experiments for these application types are provided.

\section{Background and Related Work}
\label{sec:background}

This section provides essential background information on quantum computing infrastructure, QML techniques, and related work for benchmarking.

\subsection{Quantum Computing Infrastructure}

\subsubsection*{Hardware}
Several quantum computing models exist: gate-based and analog quantum computers are the most popular models. A gate-based quantum computer controls quantum states using standardized quantum gates, such as single-qubit and two-qubit gates. Examples of gate-based devices include IBM, Google, and Rigetti's superconducting systems and ion trap systems from IonQ and Quantinuum. Analog quantum computers include D-Wave's annealing system and Pasqal's and QuEra's neutral atom platforms.

\subsubsection*{Simulation}
Simulating quantum systems is crucial as access to quantum devices is still limited. Moreover, current quantum computers have limitations regarding the number of qubits and their fidelity. High-performance computing (HPC) techniques, such as GPU-accelerated state vector (e.\,g. cuQuantum~\cite{cuquantum}) and tensor network (e.\,g., Jets~\cite{Vincent_2022} and cuTensorNet~\cite{cuquantum}) simulators, enhance runtime and enable scale.

\subsubsection*{Middleware}
Another important consideration is managing hybrid, quantum-classical workloads. For example, XACC~\cite{xacc_2020} introduces a quantum-classical programming model to integrate both computing paradigms better. CUDA Quantum~\cite{cuda_quantum} is a platform for integrating classical and quantum computing devices using a common programming model similar to XACC. Increasingly, such capabilities are integrated into existing quantum platforms, e.\,g., Qiskit Runtime~\cite{johnson2022qiskit} and Braket Jobs~\cite{braket-jobs-2021} provide mechanisms to manage classical computing with quantum tasks more efficiently. These are limited to their respective cloud environments. 

\subsection{Quantum Machine Learning}
\label{sec:qml}
Machine learning (ML)~\cite{bishop2007, Goodfellow-et-al-2016} can be defined as ``the study of computer algorithms that automatically allow computer programs to improve through experience''~\cite{Mitchell97}. ML has seen rapid growth in the last decade, and its potential applications have led to hardware advancements, new algorithms, and user-friendly software. Applications include natural language processing, computer vision, and generative design.

QML has attracted attention recently~\cite{schuld_machine_2021} as a possible realization of a \emph{quantum advantage}. Generally, a quantum generative model learns the probability distribution of a training dataset as a quantum state. The unitary map that takes an initial state, e.\,g., $|0\rangle^{\otimes N}$ of an $N$-qubit system, to its final state $|G_{\boldsymbol{\theta}}\rangle$, is usually given by a variational quantum circuit. Generative learning is achieved by finding the circuit parameters $\boldsymbol{\theta}$ such that the outcomes of measurements, in the computational basis, of quantum state $|G_{\boldsymbol{\theta}}\rangle$ are distributed according to the probability distribution of the training data. These model types take advantage of the probabilistic nature of quantum mechanics to express the probability distribution of a given dataset.  QGANs and QCBMs are two popular training techniques for quantum generative models. %

\subsubsection{Quantum Generative Adversarial Networks (QGANs)} 

A generative adversarial network (GAN) comprises two parts: a generator that creates data resembling real data and a discriminator that distinguishes between real data and the data generated by the generator. Similar to classical GANs, QGANs are trained in an adversarial way~\cite{Dallaire_Demers_2018}, i.\,e., a discriminator neural network (classical or quantum) is trained to reject synthetic data samples produced by the generator quantum system. As these two components continue to engage in this competitive game, the generator learns to produce samples that are increasingly similar to the training dataset, thereby enhancing the overall quality of the generative output. 

\subsubsection{Quantum Circuit Born Machines (QCBMs)}
Instead of requiring a discriminator neural network, QCBMs are trained on the histogram of the dataset directly. Such training is more straightforward than that of a QGAN but suffers from scaling problems, as one must compute the entire histogram of the samples from the generator at every iteration of the learning process. For large dimensional datasets, this procedure is limited by the curse of dimensionality.

\subsection{Generalization Metrics}
\label{sec:generalization}
For supervised learning tasks, the difference between performance on the training and test sets is an established metric for assessing generalization. In unsupervised learning, evaluating an algorithm's generalization capabilities is an active area of research with far-reaching implications for both classical and quantum ML~\cite{sajjadi2018assessing, BORJI201941, kynkäänniemi2019improved, borji2021pros, alaa2022faithful}. 

A generative model generalizes if the model distribution resembles the ground truth; if the model distribution resembles the distribution of the training set, it memorizes (also referred to as overfitting). Recent works~\cite{https://doi.org/10.48550/arxiv.2201.08770, https://doi.org/10.48550/arxiv.2207.13645} propose sample-based generalization metrics for discrete problems, which we review here. The ground truth distribution to be learned is represented by the solution set $\mathbb{S}$, that consists of bitstrings with a fixed length and is subject to an arbitrary condition, such as having a certain number $k$ of ones. The training set $\mathbb{T}$ is a subset of $\mathbb{S}$, and the ratio of the training set size to the solution set size is given by $\alpha \texttt{=} |\mathbb{T}| / |\mathbb{S}|$. (Multi)sets with four different conditions are determined based on the multiset of generated samples $Q$:
\begin{align*}
\mathcal{G}_\mathrm{train} &= \{q \in Q \mid q \in \mathbb{T}\} \\
\mathcal{G}_\mathrm{new} &= \{q \in Q\ |\ q \notin \mathbb{T}\} \\
\mathcal{G}_\mathrm{sol} &= \{q \in Q\ |\ q \notin \mathbb{T}, q \in \mathbb{S}\} \\
g_\mathrm{sol} &= \mathrm{set}(\mathcal{G}_\mathrm{sol})
\end{align*}
All generalization metrics are based on the respective sizes of the (multi)sets above. 

The exploration
\begin{equation*}
    E = \frac{|\mathcal{G}_\mathrm{new}|}{|Q|}
\end{equation*}
quantifies the fraction of generated samples that are unseen, both noisy and valid. The precision
\begin{equation*}
    P = \frac{|\mathcal{G}_\mathrm{train}|+|\mathcal{G}_\mathrm{sol}|}{|Q|}
\end{equation*}\label{equ:precison}%
measures the model's ability to generate data points that belong to the solution set, both seen and unseen. The fidelity 
\begin{equation*}
    F=\frac{|\mathcal{G}_\mathrm{sol}|}{|\mathcal{G}_\mathrm{new}|}
\end{equation*}
represents the model's ability to generate unseen and valid samples normalized to all new samples. The rate 
\begin{equation*}
    R=\frac{|\mathcal{G}_\mathrm{sol}|}{|Q|}
\end{equation*}
describes the ability to generate unseen valid samples normalized to the total number of samples. In our experiments, we use the normalized rate $\tilde{R}\texttt{=}R/(1 \texttt{-} \alpha)$. Finally, the coverage 
\begin{equation*}
    C=\frac{|g_\mathrm{sol}|}{|\mathbb{S}|(1-\alpha)}
\end{equation*}
captures the portion of the uncovered solution set that the model could reach. The normalized coverage is given by $\tilde{C}\texttt{=}C/\overline{C}$, where
\begin{equation*}
\overline{C} = 1-\left(1-\frac{1}{|\mathbb{S}|(1-\alpha)}\right)^{|Q|(1-\alpha)}
\end{equation*}
is the expected value of the coverage for a model that was perfectly fitted to the solution set $\mathbb{S}$.

\subsection{Benchmarking Quantum Machine Learning}
\label{sec:qml_bench}

Application-centric benchmarks, such as ImageNet~\cite{imagenet} for computer vision and Glue~\cite{wang2019glue} for natural language processing, were instrumental in advancing ML by providing labeled, standardized datasets that enabled comparisons. The proposed metrics focus primarily on the quality of ML models, e.\,g., Top 1 and Top 5 accuracy in the ImageNet benchmark.
Later, the MLCommons benchmark suite~\cite{mlcommons} (previously MLPerf~\cite{reddi2020mlperf}) emerged, proposing runtime performance, solution quality, and costs as metrics. These metrics enable analysts to understand the relationship between time-to-solution and solution quality. Currently, MLCommons comprises two training and four inference benchmarks. Each benchmark has multiple tasks, e.\,g. the training benchmark utilizes eight model architectures and datasets (including ImageNet).

The first instances of benchmarks of QML models have emerged. Benedetti et al.~\cite{Benedetti_2019} introduce the qBAS score as a hardware-independent metric based on a sampling task of the bars and stripes dataset, consisting of synthetic image data, to infer the performance of shallow circuits. Hamilton et al.~\cite{Hamilton_2019} ran similar experiments optimized for low-qubit superconducting devices. West et al.~\cite{west2022benchmarking} benchmarked QML models against their classical counterparts in the context of robustness against adversarial attacks in computer vision. Wall et al.~\cite{PhysRevResearch.3.023010} explored the performance of quantum generative models using tensor network methods based on the MNIST dataset.

\section{QUARK 2.0}
\label{sec:quark2}
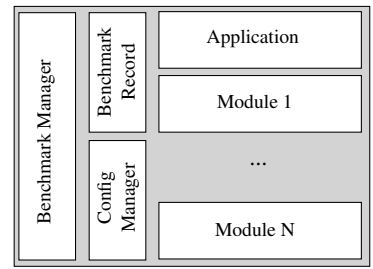
\begin{figure}[b!]
    \centering
    \resizebox{0.55\linewidth}{!}{%

\tikzset{every picture/.style={line width=0.75pt}} %

\begin{tikzpicture}[x=0.75pt,y=0.75pt,yscale=-1,xscale=1]
\draw  [fill={rgb, 255:red, 211; green, 211; blue, 211 }  ,fill opacity=1 ] (76,17) -- (415,17) -- (415,265) -- (76,265) -- cycle ;
\draw  [fill={rgb, 255:red, 255; green, 255; blue, 255 }  ,fill opacity=1 ] (214,22) -- (407,22) -- (407,76) -- (214,76) -- cycle ;
\draw  [fill={rgb, 255:red, 255; green, 255; blue, 255 }  ,fill opacity=1 ] (214,83) -- (407,83) -- (407,137) -- (214,137) -- cycle ;
\draw  [fill={rgb, 255:red, 255; green, 255; blue, 255 }  ,fill opacity=1 ] (214,204) -- (407,204) -- (407,258) -- (214,258) -- cycle ;
\draw  [fill={rgb, 255:red, 255; green, 255; blue, 255 }  ,fill opacity=1 ] (134.5,23) -- (134.5,259) -- (80.5,259) -- (80.5,23) -- cycle ;
\draw  [fill={rgb, 255:red, 255; green, 255; blue, 255 }  ,fill opacity=1 ] (201.5,23) -- (201.5,137) -- (147.5,137) -- (147.5,23) -- cycle ;
\draw  [fill={rgb, 255:red, 255; green, 255; blue, 255 }  ,fill opacity=1 ] (201.5,145) -- (201.5,259) -- (147.5,259) -- (147.5,145) -- cycle ;

\draw (258,38) node [anchor=north west][inner sep=0.75pt]  [font=\Large] [align=left] {Application};
\draw (268,100) node [anchor=north west][inner sep=0.75pt]  [font=\Large] [align=left] {Module 1};
\draw (299,165) node [anchor=north west][inner sep=0.75pt]  [font=\LARGE] [align=left] {...};
\draw (266,222) node [anchor=north west][inner sep=0.75pt]  [font=\Large] [align=left] {Module N};
\draw (96.5,226) node [anchor=north west][inner sep=0.75pt]  [font=\Large,rotate=-270] [align=left] {Benchmark Manager};
\draw (154,234) node [anchor=north west][inner sep=0.75pt]  [font=\Large,rotate=-270] [align=center] {   Config \\Manager};
\draw (154,124) node [anchor=north west][inner sep=0.75pt]  [font=\Large,rotate=-270] [align=center] {Benchmark\\Record};

\end{tikzpicture}
     }\upp
    \caption{\textbf{QUARK Architecture:} Based on the new \texttt{Core} module, we can construct benchmark workflows with an arbitrary number of concrete modules such as \texttt{Module 1}, always starting with an application but not limited anymore to the previous four required module types. } \upp\upp
    \label{fig:architecture}
\end{figure}

QUARK 2.0 refines the original QUARK architecture to meet the complex requirements of QML benchmarks. The original architecture's design focused on optimization tasks and could not appropriately accommodate multiple training and test datasets, data transformations, model architectures, hyperparameters, and QML metrics. 
By revising the architecture, we remove these limitations and allow the development of complex benchmark workflows while still supporting the original applications, such as the polyvinyl chloride (PVC) sealing process and the traveling salesperson problem (TSP).

\subsection{Architecture}
In QUARK\,2.0, we generalize the architecture and abstractions. Instead of four predefined components, we introduce the abstract base class \texttt{Core}, which defines every module's mandatory attributes and functions.
Using this Core module, we provide the implementation guidelines for all other concrete or abstract realizations of this module. This ensures a standardized interface, enforces the structure needed to execute a benchmark, and thus enables the design of a benchmark with an arbitrary number of modules, as depicted in Figure~\ref{fig:architecture}. 

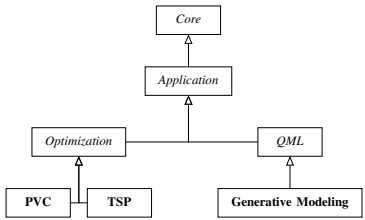
\begin{figure}[b!]
    \centering
    \resizebox{0.55\linewidth}{!}{%
\tikzumlset{fill class = white, fill template = white}
\begin{tikzpicture}

\umlsimpleclass[type = abstract, x=0, y=0]{Core}

\umlsimpleclass[type = abstract, x=0, y=-1.5 ]{Application}
\umlsimpleclass[type = abstract, x=-2.7, y=-3 ]{Optimization}
\umlsimpleclass[type = abstract, x=2.5, y=-3 ]{QML}
\umlsimpleclass[x=-1.7, y=-4.5 ]{TSP}
\umlsimpleclass[x=-3.7, y=-4.5 ]{PVC}
\umlsimpleclass[x=2.5, y=-4.5 ]{Generative Modeling}
\umlinherit[geometry=-|]{Application}{Core}
\umlinherit[geometry=-|]{Optimization}{Application}
\umlinherit[geometry=-|]{QML}{Application}
\umlinherit[geometry=-|]{TSP}{Optimization}
\umlinherit[geometry=-|]{PVC}{Optimization}
\umlinherit[geometry=-|]{Generative Modeling}{QML}

\end{tikzpicture}     }\upp
    \caption{An example of how all modules are based on the \texttt{Core} module and can be further specified in abstract implementations such as the \texttt{QML} module.} \upp\upp
    \label{fig:uml}
\end{figure}

The only mandatory subclass of the \texttt{Core} module, which has to be the first module of every benchmark workflow, is an implementation of an \texttt{Application} module.

To make the \texttt{Application} module more general, we created a new abstract \texttt{Optimization} class, which extends the \texttt{Application} class by the required methods for optimization problems, which were previously specified in the \texttt{Application} module. This way, users can create abstract classes based on the \texttt{Core} module to describe the requirements for their module types, as depicted in Figure~\ref{fig:uml}.

The essential functions of the \texttt{Core} module, which can be used by every module/subclass to execute its logic (see Figure~\ref{fig:benchmark_process}), are the \texttt{preprocess} and the \texttt{postprocess} method. The \texttt{preprocess} method executes before the input data is passed to the subsequent module while the \texttt{postprocess} method executes before the data is passed back to the preceding module. If no subsequent or preceding module exists, these functions are still executed. The interplay between \texttt{pre-} and \texttt{postprocess} can be exemplified through data transformation for a QML application. The data is transformed into the training space during preprocessing, while \texttt{postprocess} applies the inverse transformation.

The new \texttt{Config Manager} encapsulates the generation and instantiation of the benchmark configuration, which the \texttt{Benchmark Manager} then uses to execute the actual benchmark. 

\subsubsection*{Data Collection}
For each benchmark run, we also instantiate a single \texttt{Benchmark Record}, which contains, alongside some general information about the benchmark run (e.\,g., the git revisions of the framework), it's metadata, such as the configuration, and metrics from all the modules in a benchmark workflow. 
This way, metrics collection across all module levels can be collected, allowing the user to add metrics at every module level, enabling more advanced analysis of a benchmark run.

These metrics also include the time executed by the \texttt{preprocess} and the \texttt{postprocess}, $T_{\mathrm{preprocess}}$ and $T_{\mathrm{postprocess}}$ respectively. When summed up, these two metrics combine to the total time spent in a specific module $T_{\mathrm{module \ N}}$:

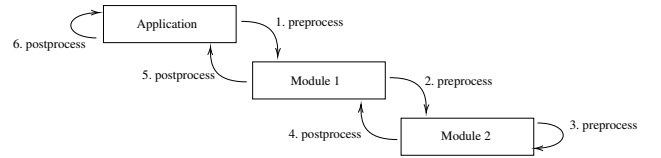
\begin{figure}[b!]
    \centering
    \resizebox{0.95\linewidth}{!}{%

\tikzset{every picture/.style={line width=0.75pt}} %

\begin{tikzpicture}[x=0.75pt,y=0.75pt,yscale=-1,xscale=1]
\draw  [fill={rgb, 255:red, 255; green, 255; blue, 255 }  ,fill opacity=1 ] (142,81) -- (335,81) -- (335,135) -- (142,135) -- cycle ;
\draw  [fill={rgb, 255:red, 255; green, 255; blue, 255 }  ,fill opacity=1 ] (359,163) -- (552,163) -- (552,217) -- (359,217) -- cycle ;
\draw  [fill={rgb, 255:red, 255; green, 255; blue, 255 }  ,fill opacity=1 ] (575,245) -- (768,245) -- (768,299) -- (575,299) -- cycle ;
\draw [line width=0.75]    (342.5,104) .. controls (390.52,104) and (396.28,127.05) .. (395.55,153.38) ;
\draw [shift={(395.5,155)}, rotate = 272.12] [color={rgb, 255:red, 0; green, 0; blue, 0 }  ][line width=0.75]    (12.02,-3.62) .. controls (7.65,-1.54) and (3.64,-0.33) .. (0,0) .. controls (3.64,0.33) and (7.65,1.54) .. (12.02,3.62)   ;
\draw [line width=0.75]    (558.5,186) .. controls (606.52,186) and (612.28,209.05) .. (611.55,235.38) ;
\draw [shift={(611.5,237)}, rotate = 272.12] [color={rgb, 255:red, 0; green, 0; blue, 0 }  ][line width=0.75]    (12.02,-3.62) .. controls (7.65,-1.54) and (3.64,-0.33) .. (0,0) .. controls (3.64,0.33) and (7.65,1.54) .. (12.02,3.62)   ;
\draw [line width=0.75]    (569.48,276.13) .. controls (521.46,275.89) and (515.81,252.81) .. (516.68,226.48) ;
\draw [shift={(516.74,224.87)}, rotate = 92.41] [color={rgb, 255:red, 0; green, 0; blue, 0 }  ][line width=0.75]    (12.02,-3.62) .. controls (7.65,-1.54) and (3.64,-0.33) .. (0,0) .. controls (3.64,0.33) and (7.65,1.54) .. (12.02,3.62)   ;
\draw [line width=0.75]    (350.48,192.13) .. controls (302.46,191.89) and (296.81,168.81) .. (297.68,142.48) ;
\draw [shift={(297.74,140.87)}, rotate = 92.41] [color={rgb, 255:red, 0; green, 0; blue, 0 }  ][line width=0.75]    (12.02,-3.62) .. controls (7.65,-1.54) and (3.64,-0.33) .. (0,0) .. controls (3.64,0.33) and (7.65,1.54) .. (12.02,3.62)   ;
\draw [line width=0.75]    (773.5,252) .. controls (822.01,252) and (825.44,287.28) .. (771.16,288.96) ;
\draw [shift={(769.5,289)}, rotate = 358.98] [color={rgb, 255:red, 0; green, 0; blue, 0 }  ][line width=0.75]    (12.02,-3.62) .. controls (7.65,-1.54) and (3.64,-0.33) .. (0,0) .. controls (3.64,0.33) and (7.65,1.54) .. (12.02,3.62)   ;
\draw [line width=0.75]    (131.69,128.17) .. controls (83.18,127.68) and (80.11,92.36) .. (134.4,91.23) ;
\draw [shift={(136.06,91.21)}, rotate = 179.56] [color={rgb, 255:red, 0; green, 0; blue, 0 }  ][line width=0.75]    (12.02,-3.62) .. controls (7.65,-1.54) and (3.64,-0.33) .. (0,0) .. controls (3.64,0.33) and (7.65,1.54) .. (12.02,3.62)   ;

\draw (189,99) node [anchor=north west][inner sep=0.75pt]  [font=\Large] [align=left] {Application};
\draw (413,182) node [anchor=north west][inner sep=0.75pt]  [font=\Large] [align=left] {Module 1};
\draw (632,263) node [anchor=north west][inner sep=0.75pt]  [font=\Large] [align=left] {Module 2};
\draw (391,97) node [anchor=north west][inner sep=0.75pt]   [align=left, font=\Large] {1. preprocess};
\draw (608,183) node [anchor=north west][inner sep=0.75pt]   [align=left, font=\Large] {2. preprocess};
\draw (819,247) node [anchor=north west][inner sep=0.75pt]   [align=left, font=\Large] {3. preprocess};
\draw (410,259) node [anchor=north west][inner sep=0.75pt]   [align=left, font=\Large] {4. postprocess};
\draw (197,174) node [anchor=north west][inner sep=0.75pt]   [align=left, font=\Large] {5. postprocess};
\draw (9,129) node [anchor=north west][inner sep=0.75pt]   [align=left, font=\Large] {6. postprocess};

\end{tikzpicture}     %
    }\upp
    \caption{\textbf{Benchmark Process:} Visualization of how the benchmark process is designed and how the input is preprocessed and recursively passed down in the chain of modules until the end is reached. Then, after a postprocessing step, the outputs of the modules are passed up the chain until the head of the chain (always an \texttt{Application}) is reached. There the benchmark process ends.}\upp\upp
    \label{fig:benchmark_process}
\end{figure}

\begin{equation}
\label{eq:metrics}
\begin{aligned}
T_{\mathrm{module \ N}} ={} & T_{\mathrm{preprocess}} + T_{\mathrm{postprocess}}.
\end{aligned}
\end{equation}

$TTS$ is defined as the end-to-end time required to run a benchmark. It is decomposed into several components:
\begin{equation}
\label{eq:tts}
\begin{aligned}
TTS ={} & T_{\mathrm{application}} + T_{\mathrm{module \ 1}} + ... + T_{\mathrm{module \ N}}.
\end{aligned}
\end{equation}

At the end of a QUARK run, the data collected in the various \texttt{Benchmark Record} objects are extracted and stored in a single JSON file, accommodating easy analysis.

With these architectural changes, we can now use the framework for applications not just from the area of optimization but also from fields like QML, which require the design of various new module definitions. 
However, it is essential to note that while these enhancements change the existing architecture of QUARK, we still adhere to its original design principles by providing reproducibility, verifiability, high usability, customizability, and easy collection of benchmark results.

\subsection{Application: Quantum Generative Modeling}
\begin{figure}[t!]
    \centering
    \resizebox{\linewidth}{!}{%
     \includegraphics[width=\textwidth]{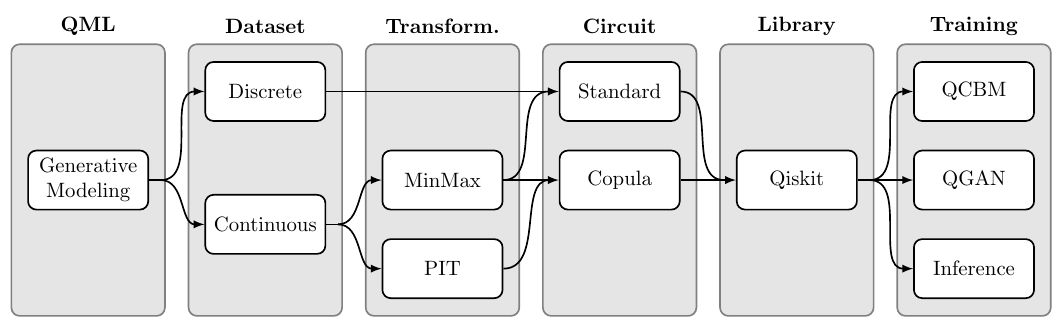}
    }\upp
    \caption{\textbf{The Generative Model} embedded in QUARK\,2.0. The six layers (i.\,e., Application, Dataset, Transformation, Circuit, Library, Training) represent abstract base classes for QML applications. The white nodes indicate the concrete implementation, such as the arrangements of quantum gates in a quantum circuit.}
    \label{fig:architecture_QML}
\end{figure}

Benchmarking quantum generative modeling involves decomposing the workflow into modular components, designing representative workloads, and measuring performance. The decomposition enables extension, scalability, and comparisons between different systems. The components are \texttt{QML}, \texttt{Dataset}, \texttt{Transformation}, \texttt{Circuit}, \texttt{Library} and \texttt{Training} (see Figure~\ref{fig:architecture_QML}). The benchmark is characterized by the configuration file, enabling the reproducibility of experiments and facilitating the execution of new instances with varying features, thus enabling easy performance comparisons.

\subsubsection{Dataset}
The QUARK framework offers the flexibility of selecting many synthetic and real datasets. The \texttt{Discrete} module generates cardinality-constrained probability distributions as described by Gili at al.~\cite{https://doi.org/10.48550/arxiv.2207.13645}. The solution space $\mathbb{S}$ is defined by the bitstrings with a Hamming weight (i.\,e. number of ones) of $k$. Each digit in the bitstring will correspond to a measured value of a qubit. 

The \texttt{Continuous} module includes low-dimensional synthetic and real datasets as described by Riofr\'io et al.~\cite{riofrio2023performance}, including mixed Gaussians, datasets resembling the shape of the letter X and O, and a time series of stock prices from Yahoo! Finance~\cite{yfinance}. 

\subsubsection{Transformation}
As described above, the continuous datasets are passed on to either the \texttt{MinMax} normalization or the probability integral transformation \texttt{PIT}. The former maps the $m$ marginal distributions of the dataset to the interval $[0, 1]$ by applying an affine transformation. The latter transforms the original probability distribution to the distribution of its uniformly distributed cumulative marginals, known as the copula~\cite{doi:10.1137/1036146}. Preprocessing applies the transformation, while postprocessing applies its inverse.

\begin{figure}[t!]
    \centering
    \resizebox{\linewidth}{!}{%
     \includegraphics[width=\textwidth]{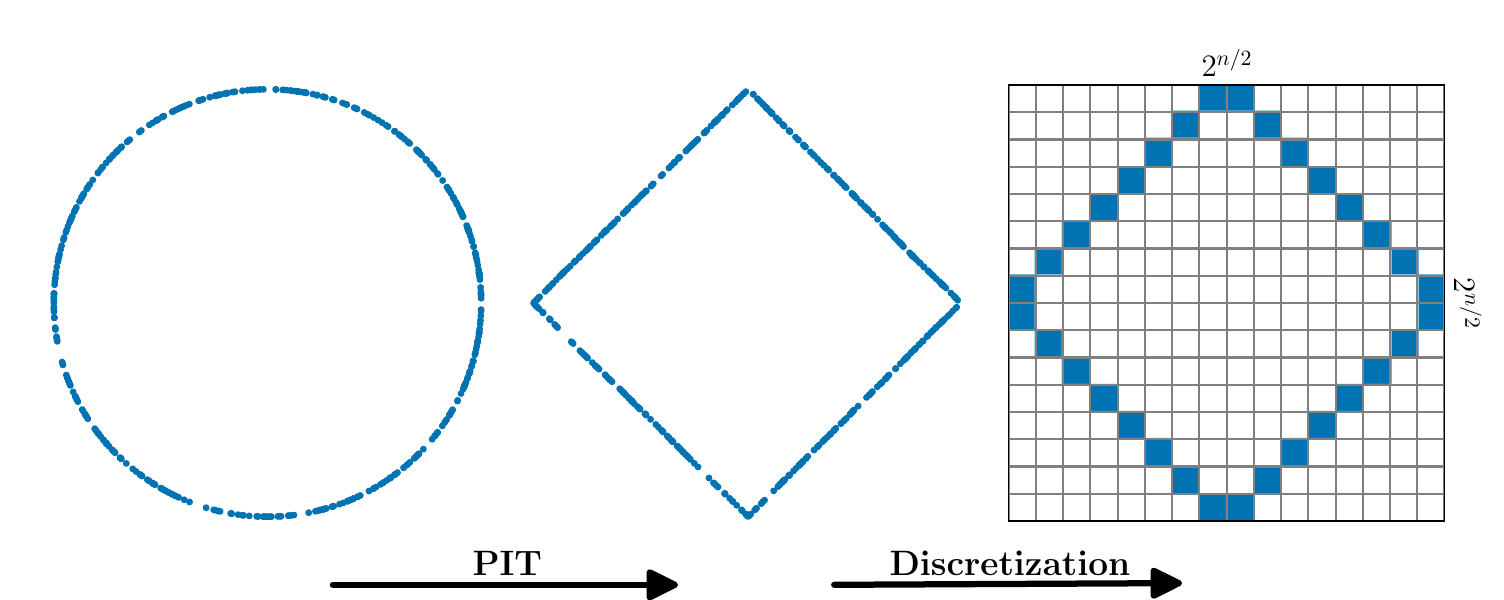}
    }\upp
    \caption{\textbf{Transformation} of the dataset resembling the letter O (left) in two stages: probability integral transformation (middle) and subsequent discretization (right). The size of the discretization grid is given by ($2^{n/2} \times 2^{n/2}$), with the number of qubits $n$.}
    \label{fig:dataset}
\end{figure}

The transformed dataset is mapped to a discrete grid, and its discrete probability distribution, called probability mass function (PMF), is determined. The granularity of the grid corresponds to the number of basis states in the search space $|\mathcal{U}|\texttt{=}2^n$. At this stage, both continuous and discrete datasets are represented by their PMFs. Figure~\ref{fig:dataset} depicts the transformation of the dataset resembling the letter O.

\subsubsection{Circuit}
\begin{figure*}[!t]
    \centering
    \resizebox{\linewidth}{!}{
     \subfloat[]{\includegraphics[width=0.51\textwidth]{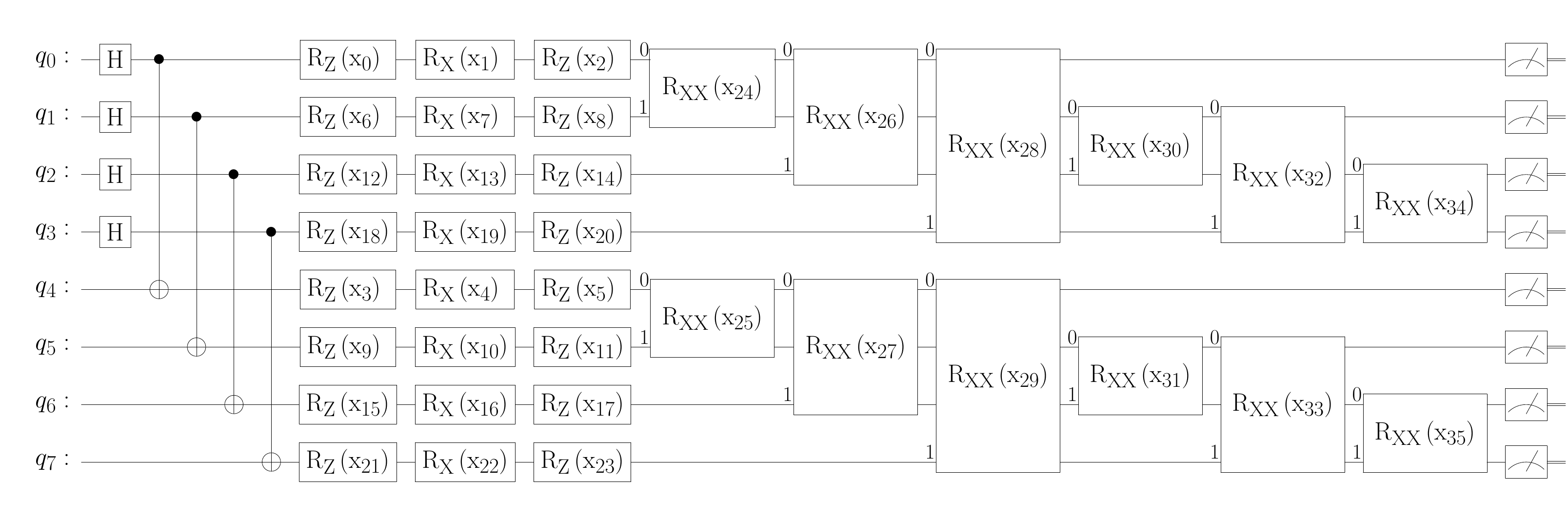}}
     \subfloat[]{\includegraphics[width=0.47\textwidth]{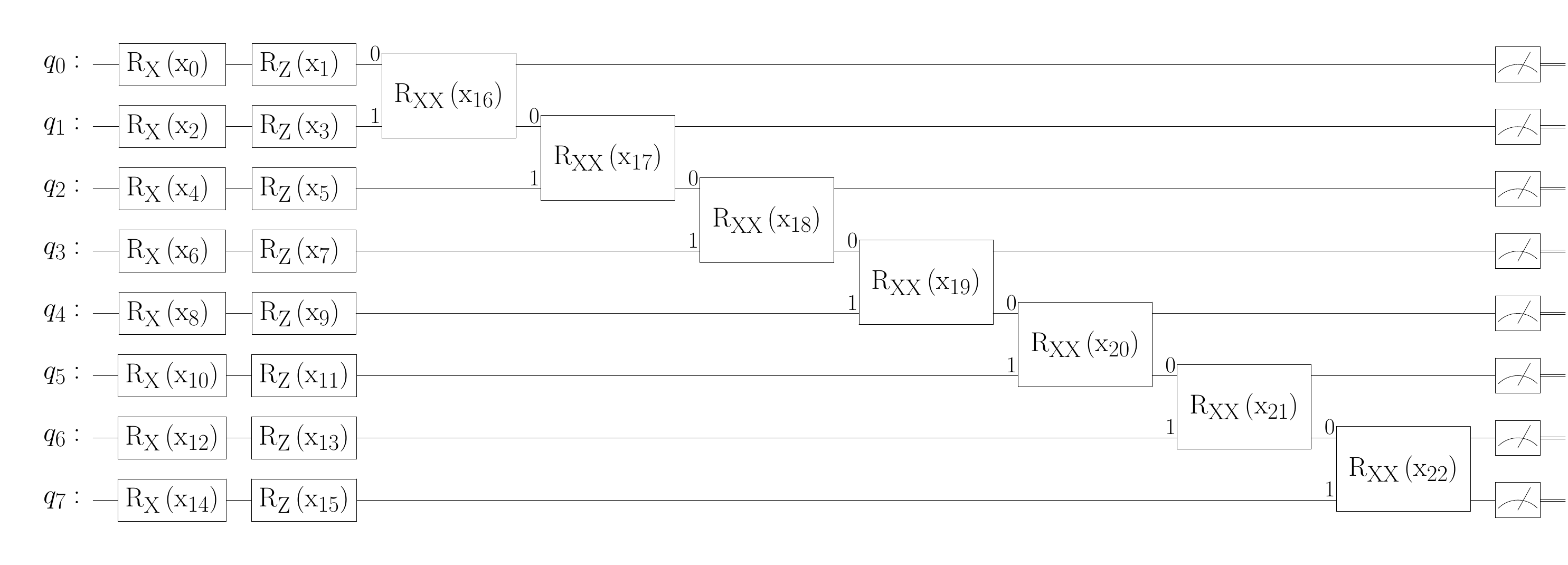}}}
    \caption{The (a) \textbf{copula circuit} and (b) \textbf{standard circuit}. The parameter count of the copula circuit is given by $3n \texttt{+} m \cdot \binom{n/2}{m}$, with the number of qubits $n$ and the number of registers $m$ (i.\,e. the number of marginal distributions). The parameter count of the standard circuit for an even number of layers is given by $3n\texttt{-}1$ for $d\texttt{=}2$ and $(3d/2\texttt{+}1)n\texttt{-}(d/2)$ for $d$$>$$2$, with the circuit depth $d$.}\upp\upp
    \label{fig:circuit}
\end{figure*}

After selecting the \texttt{Standard} or \texttt{Copula} circuit and specifying the number of qubits and the circuit depth, the corresponding gate sequence is returned, comprised of a list of quantum gates and the associated wires. Scaling the quantum circuits by varying the number of qubits and circuit depth enables adjusting the workload. The copula architecture is optimized to learn the probability distribution whose cumulative marginals are uniformly distributed like the image of the PIT~\cite{Zhu_2022}.  Figure~\ref{fig:circuit} depicts the quantum circuits.

\subsubsection{Library}
The \texttt{Library} module maps the gate sequence to the library-specific definition of a quantum circuit and wraps the circuit in the \texttt{execute circuit} function. This function returns the measurement outcome for a given set of model parameters and is passed to the \texttt{Training} module. 

The library-agnostic implementation of the gate sequence and training modules will simplify the extension of QUARK with libraries, such as PennyLane~\cite{bergholm2022pennylane} and Cirq~\cite{cirq_developers_2022_7465577}.

\subsubsection{Training}
After the hyperparameters of the training are selected by reading the configuration file, %
the \texttt{train} function orchestrates the dataset transformation, circuit execution, and optimization loop. The selected circuit ansatz is parametrized by free variational parameters $\boldsymbol{\theta}$. Then, the variational parameters of the \texttt{Circuit} are adjusted to fit the model distribution to the transformed \texttt{Dataset}. Two training methods are supported, the \texttt{QGAN} and \texttt{QCBM}.

The training loop can be made either by gradient descent methods based on the parameter shift rule (for QGAN) or covariance matrix adaptation evolutionary strategy CMA-ES~\cite{nikolaus_hansen_2023_7573532} (for QCBM), which is a gradient-free method, that models the correlations among candidate solutions as Gaussian variables based on their loss function value. Both algorithms iteratively update the variational quantum circuit parameters to minimize an overall loss function. The loss function used for our studies is the Kullback-Leibler (KL) divergence, 
\begin{equation}
    \label{equ:KL_divergence}
    C_\mathrm{KL}(p_\mathrm{target}, p_\mathrm{model}) = \sum_{i=0}^{N_d-1}p_{\mathrm{target},i}\ln{\frac{p_{\mathrm{target}, i}}{ p_{\mathrm{model},i}}},
\end{equation}
with $N_d$ defining the discretization grid in which the probability distributions are given as a measure of distance between the probability distribution of the data ($p_\mathrm{Target}$) and that learned by the quantum generator ($p_\mathrm{Model}$).
For the model's training, $N_d$ is set to $2^n$ with the number of qubits $n$.

Finally, when the model parameters, which minimize the loss function, are found, we can use the trained model to generate synthetic data. For this purpose, QUARK's \texttt{Inference} module can produces samples from the probability distribution learned during training.

\section{Performance Characterization}
\label{sec:experiments}

We demonstrate the capabilities of QUARK\,2.0 by running an extensive characterization of the quantum generative model application presented~\ref{sec:quark2}. The results aim not to highlight the best approach to solve a given problem but demonstrate the capabilities of QUARK\,2.0 to support QML workflows.

\subsection{Experimental Design}
All presented results are based on generative quantum models fitted to the datasets by minimizing the KL divergence between the output and target distribution with the CMA-ES optimizer. We (1) investigate the training performance for different hyperparameter configurations, (2) compare the performance of different state vector simulators, (3) deploy pre-trained models on IonQ Harmony, and (4) study the generalization capabilities of QCBMs. 

For experiments (1-3), we use the copula circuit with the synthetic X dataset and 10,000 samples to estimate the model distribution. To compare the experimental results, we calculate the KL divergences between the model and target distribution using a resolution corresponding to a discretization grid of ten qubits (i.e., 1024 bins). In Experiment (4), we utilize the standard circuit with twelve qubits and the cardinality-constrained dataset, employing the precise probabilities from the state vector simulator as the model distribution. 

Runtime experiments with quantum simulators were conducted on an \emph{NVIDIA DGX A100} machine (Dual AMD EPYC 7742, 2 TB memory, 8x NVIDIA A100 40 GB) using the NVIDIA cuQuantum Appliance 22.11~\cite{cuqauntum2211}, a software development kit for accelerating quantum computing workflows by distributing state vector simulations over multiple GPUs.

The experiments on IonQ Harmony were conducted using Amazon Braket Hybrid Jobs.

\subsection{Training Performance}
\label{sec:training_performance}

\begin{figure}[b!]
    \centering
    \resizebox{\linewidth}{!}{%
     \includegraphics[width=\textwidth]{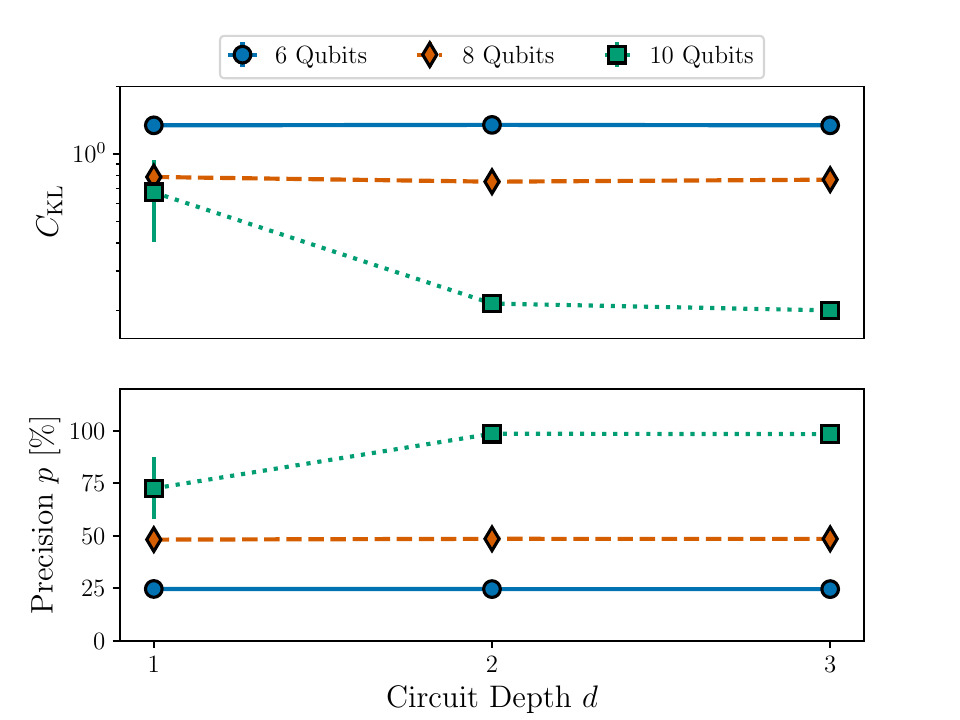}
    }\upp
    \caption{\textbf{KL Divergence} (top) and \textbf{Precision} (bottom) in dependence on the circuit depth for the X dataset for different numbers of qubits. The KL divergence, computed in a $32 \times 32$ grid for all cases, is averaged across five experimental runs, with the error bars given by the standard error on the mean ($\sigma / \sqrt{5}$).}
    \label{fig:parameter_studies}
\end{figure}

We fitted the copula circuit with the QCBM to the X dataset, varying the number of qubits $n$$\in$$\{6, 8, 10\}$ and circuit depth $d$$\in$$\{1,2,3\}$ as presented in Figure~\ref{fig:parameter_studies}. All configurations were trained with CMA-ES, with a population size of 200 and a stopping criterion of 20,000 evaluations. The KL divergence in dependence on the epoch was investigated to ensure that the loss function converged.

With increasing model parameters, fitting the model distribution to the target distribution becomes more challenging. Therefore, model parameters of previously fitted shallow circuits were used to initialize deeper circuits, e.\,g., by initializing the parameters of the first block of the circuit with $d\texttt{=}2$ with the parameters of the trained model with $d\texttt{=}1$. The subsequent blocks were initialized with parameters close to zero. The closer the values of the parameters are to zero, the closer the resulting unitary will be to the identity operation, and the structure of the lower-depth circuit is preserved.

The KL divergence and precision $P$ were computed between the model output distribution of models with six, eight, and ten qubits and the target distribution of the model with ten qubits. $P$ is given by the ratio of valid generated samples to the total number of generated samples. As the resolution during training increases with the number of qubits, larger models show better performance (i.\,e., lower KL divergence and higher precision).

Despite the model with ten qubits, increasing the circuit depths did not improve the model's performance. The large standard error on the mean (SEM) for the KL divergence and precision of the ten qubit circuit with a depth of one suggests that the observed performance improvement is likely attributed to an outlier.

\subsection{GPU Acceleration}
\label{sec:acceleration}
\begin{figure}[t!]
    \centering
    \resizebox{\linewidth}{!}{%
     \includegraphics[width=\textwidth]{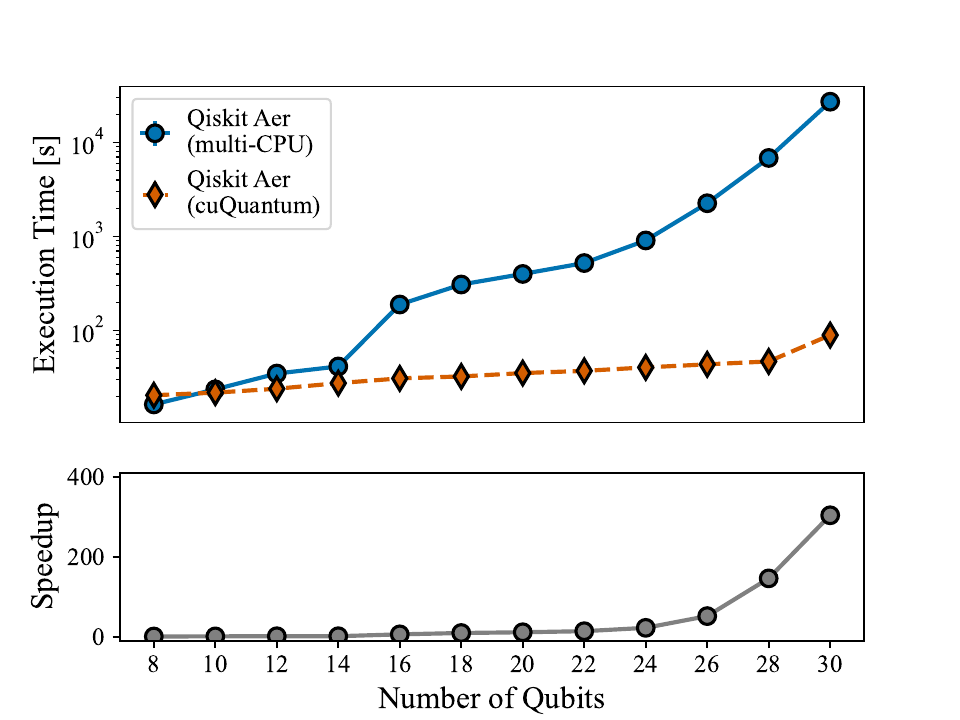}
    }\upp
    \caption{\textbf{Circuit Execution Time} of the copula circuit with a depth of two, leveraging Qiskit AerSimulator with the cuQuantum multi-GPU and the standard multi-CPU backend. The results are averaged across three iterations in the training loop. The standard error on the mean ($\sigma / \sqrt{5}$) is negligible. The maximal speed-up is reached for 30 qubits with a value of 300.}\upp\upp
    \label{fig:gpu_performance}
\end{figure}

We evaluate the performance of the Qiskit Aer state vector simulator (AerSimulator) by comparing the NVIDIA cuQuantum and multi-CPU backend. Both simulators used the default configuration with single-precision floating-point math (FP32). For the cuQuantum simulation, we distribute the workload over eight GPUs via separate MPI processes. For the CPU simulation, we use a single core for up to 14 qubits, and all CPU cores (i.\,e. 2\,CPUs, with 64\,Cores/128\,Threads each) for larger circuits as specified in the default Qiskit AerSimulator configuration.

Figure~\ref{fig:gpu_performance} shows the execution time dependent on the number of qubits for the Qiskit AerSimulator with cuQuantum and multi-CPU backend. The copula circuit was executed in each experiment with 200 independent sets of model parameters. The maximum speed-up is reached for 30 qubits, with a value of 300. The threshold value of 14 of the default Qiskit AerSimulator results in the discontinuity of the execution time for multi-CPU experiments.

As the scale of quantum systems increases, the bottlenecks in classical parts of hybrid algorithms become more apparent. For $n$ qubits, the KL divergence must be computed between all $N_d\texttt{=}2^n$ bins of the model and target distribution, translating to roughly $10^9$ function calls for 30 qubits. Therefore, the sparsity of the target and model distributions was used to accelerate the computation. The number of non-zero bins in the target distribution of the X dataset is given by $2^{1 \texttt{+} n \texttt{/} 2}$. The number of non-zero bins of the model distribution is bounded from above by the shot number. The KL divergence is computed only for bins where at least the model or target distribution is non-zero. Furthermore, the processing was accelerated with cupy~\cite{cupy_learningsys2017}. The computation time of the KL divergence ranges from $0.8$ sec for eight qubits to $1.3$ sec for 30 qubits and is not included in the circuit execution times in Figure~\ref{fig:gpu_performance}.

\subsection{Deployment on Quantum Hardware}
\label{sec:hardware}
To assess the impact of noise on quantum hardware, we compared the KL divergence of the target distribution and samples generated using a pre-trained model on two different platforms: the noise-free Qiskit AerSimulator and IonQ Harmony.

\begin{figure}[t!]
    \centering
    \resizebox{\linewidth}{!}{%
     \includegraphics[width=\textwidth]{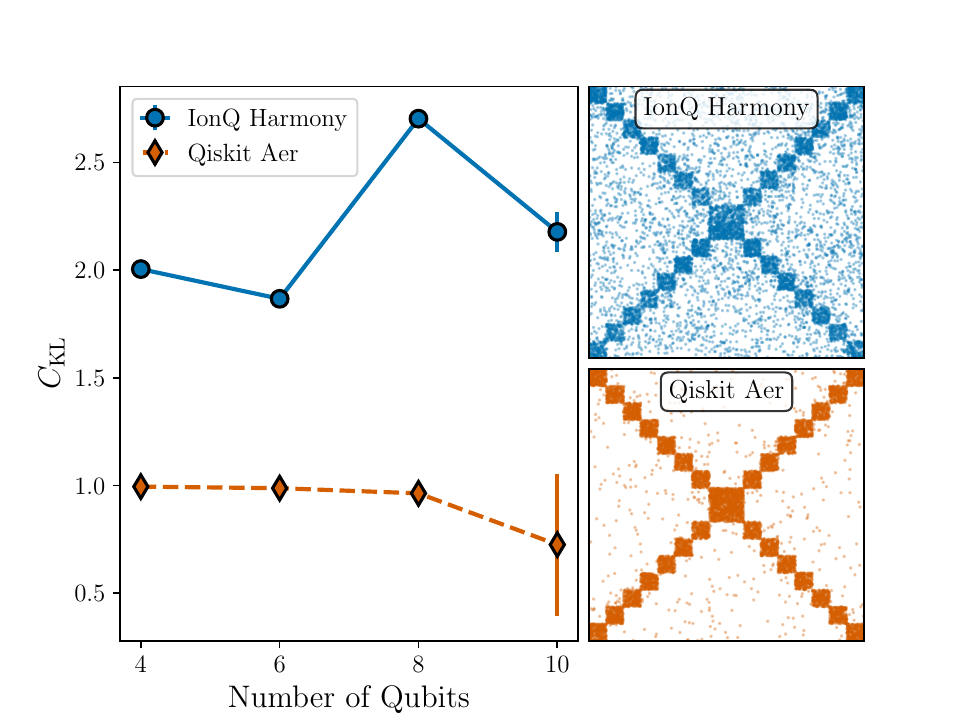}
    }\upp
    \caption{\textbf{KL Divergence} (left) in dependence on the number of qubits for the noise-free Qiskit AerSimulator and IonQ Harmony. Each experiment is based on the copula circuit with parameters of a model that was previously fitted to the dataset resembling the shape of the letter X. The results are averaged across five experiments, and the error bars indicate the standard error on the mean ($\sigma/\sqrt{5}$). Furthermore, the underlying sample distributions (right) of an eight-qubit experiment for both Qiskit AerSimulator (orange) and IonQ Harmony (blue) are shown. The eight qubits translate to a grid size of $2^8 = 256$. For IonQ Harmony, the influence of quantum errors blurs the sample distribution.}
    \label{fig:hardware}
\end{figure}

As shown in Figure~\ref{fig:hardware}, the KL divergence for both platforms shows no correlation with the number of qubits. Due to noise, the KL divergence for IonQ Harmony is higher than for the Qiskit AerSimulator. For IonQ Harmony, the KL divergence fluctuates significantly beyond the SEM due to variations in the calibration of the quantum device. 

\begin{figure*}[t!]
\captionsetup[subfloat]{farskip=0pt,margin={0.55cm,0cm}}
    \centering
     \subfloat[]{\includegraphics[width=0.32\textwidth]{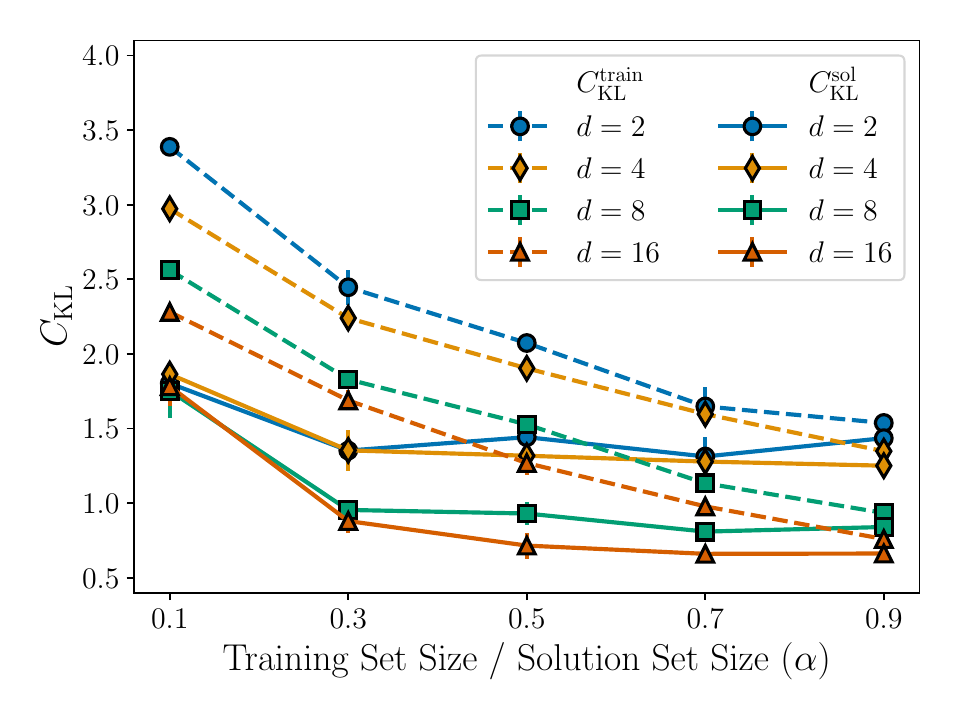}}
     \subfloat[]{\includegraphics[width=0.32\textwidth]{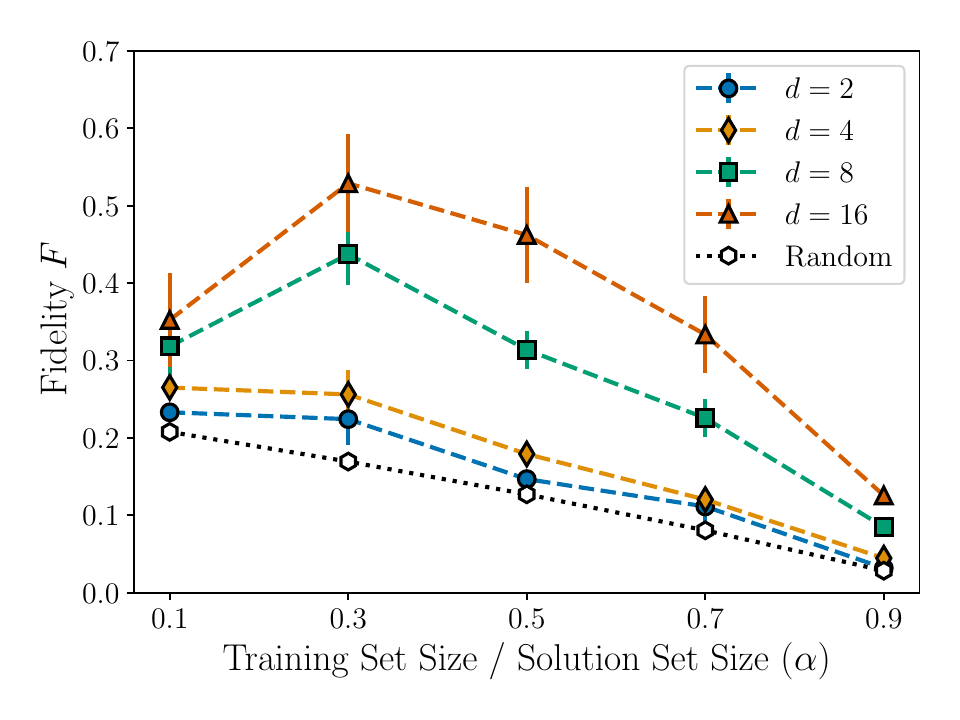}}
     \subfloat[]{\includegraphics[width=0.32\textwidth]{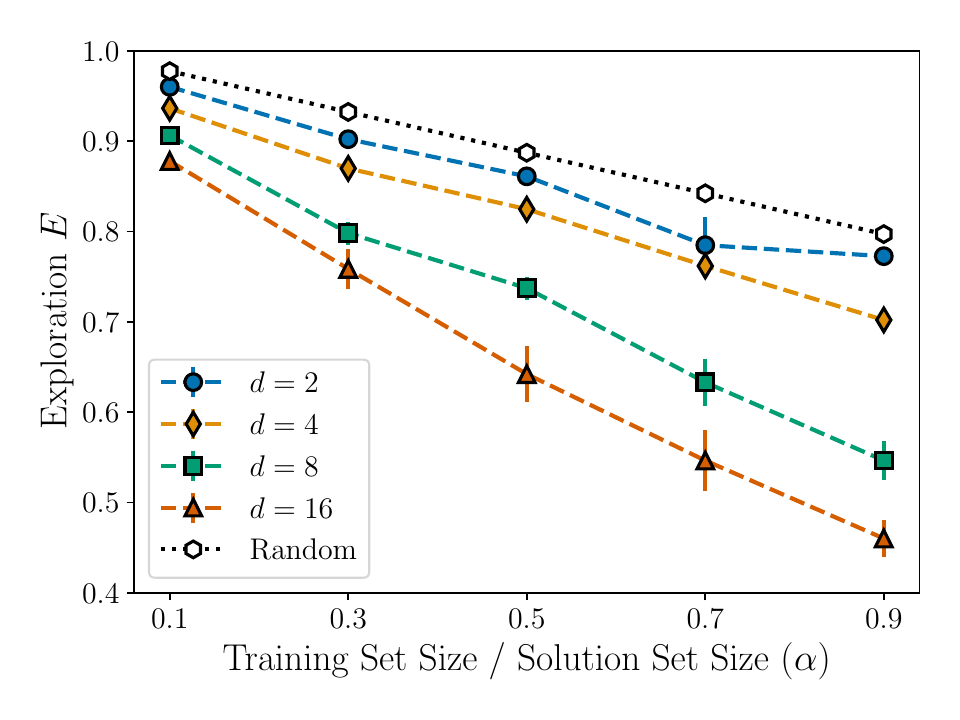}}
     
     \subfloat[]{\includegraphics[width=0.32\textwidth]{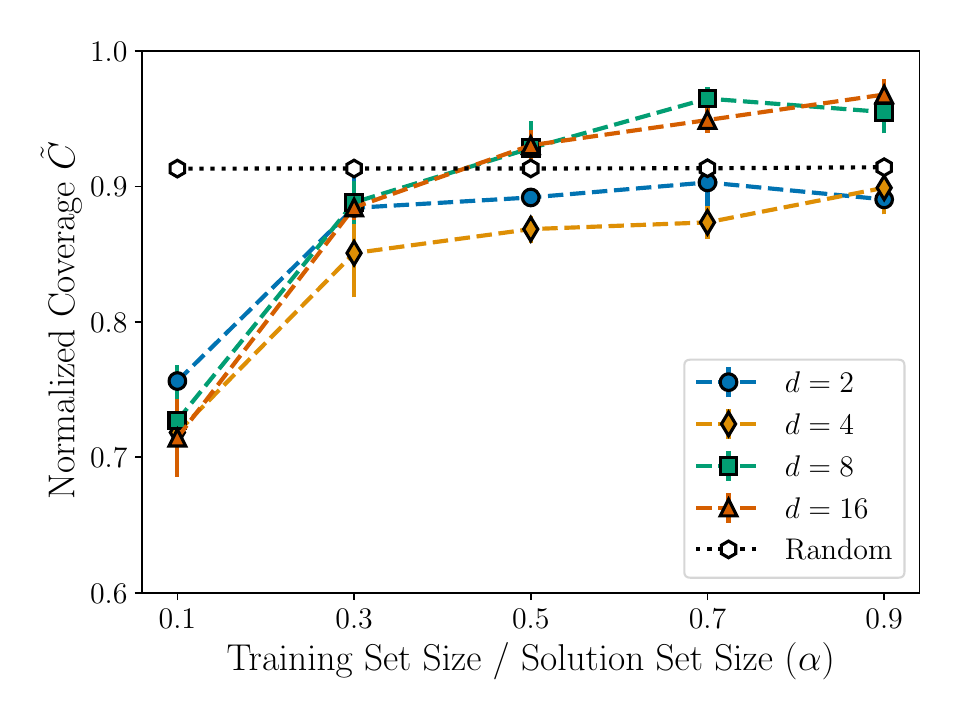}}
     \subfloat[]{\includegraphics[width=0.32\textwidth]{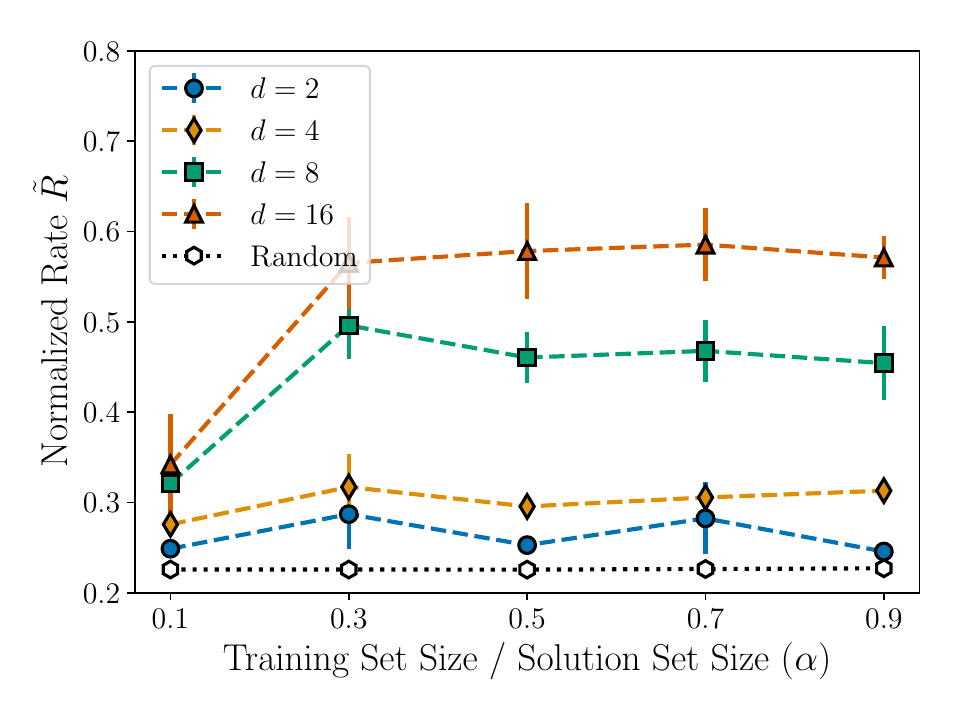}}
     \subfloat[]{\includegraphics[width=0.32\textwidth]{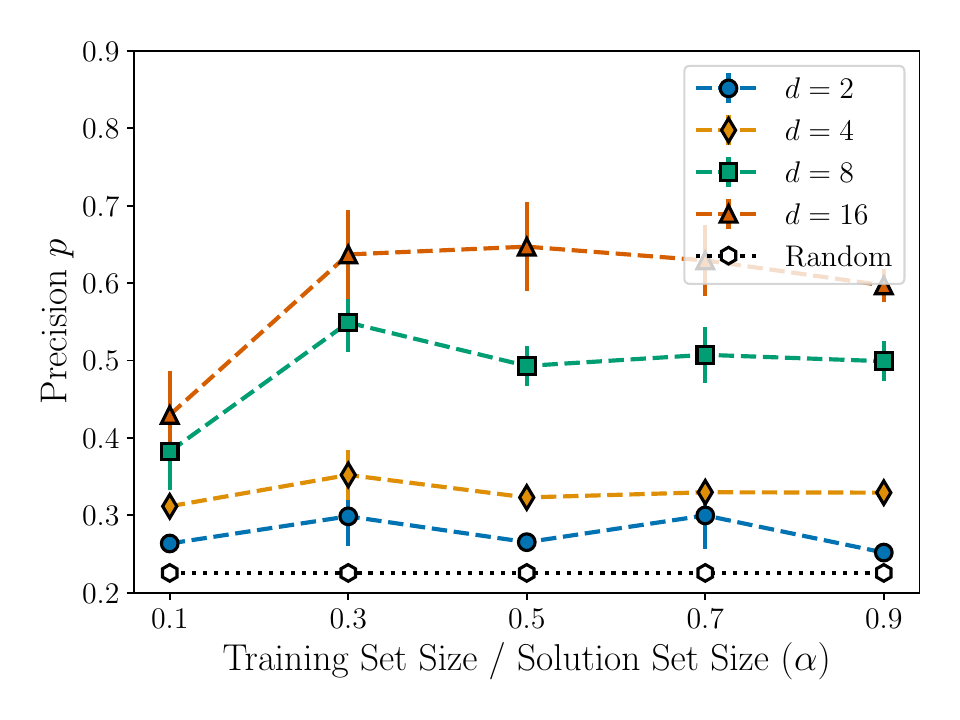}}
    \caption{\textbf{Generalization Performance of the QCBM} across various ratios of training set size to solution set sizes $\alpha$. The (a) KL divergence, (b) fidelity, (c) exploration, (d) normalized coverage, (e) normalized rate, and (f) precision are presented in dependence on the fraction of the solution set used for training $\alpha$$\in$$\{0.1, 0.3, 0.5, 0.7, 0.9\}$ for circuit depths $d$$\in$$\{2, 4, 8, 16\}$. The results are averaged across five experiments, and the error bars indicate the standard error on the mean ($\sigma/\sqrt{5}$). The best generalization is achieved for a depth of $d\texttt{=}12$ and $\alpha \texttt{=} 0.5$ with a fidelity and normalized rate of $0.52\pm0.04$ and $0.63\pm0.04$, respectively.}
    \label{fig:experiments_generalisation}
\end{figure*}

\subsection{Generalization}
\label{sec:gneralization}
In the following, we replicate the studies conducted by Gili et al.~\cite{https://doi.org/10.48550/arxiv.2207.13645} about the generalization performance toward unseen and valid samples of QCBMs. Therefore, we trained the QCBM with twelve qubits on different fractions $\alpha$ of the solution set $\mathbb{S}$ and different circuit depths $d$. As we increased the circuit size, it became evident that QCBMs are susceptible to the curse of dimensionality, as the uncertainty in the model distribution, estimated from 10,000 shots, was too high. Therefore, we used the measurement probability vector of the statevector simulator as the model distribution for the generalization experiments. For the different configurations, we determined the KL divergence, fidelity, exploration, normalized coverage, normalized rate and precision (see Figure~\ref{fig:experiments_generalisation}). The dataset consists of bitstrings with a length of twelve, corresponding to the number of qubits. The solution set consists of all bitstrings that have precisely $k\texttt{=}6$ ones (e.\,g., 010110100110 is valid, 011001000001 is not valid), resulting in a solution set size of $|\mathbb{S}| \texttt{=} 924$. 

Increasing the circuit depth $d$ improves the model's performance across all evaluated metrics. Moreover, the model outperforms the random baseline with deeper circuits while showing marginal improvement for shallower ones. The KL divergence decreases with increasing $\alpha$. We compute the KL divergences $C_\mathrm{KL}^\mathrm{train}$ and $C_\mathrm{KL}^\mathrm{sol}$ between the model distribution and training set $\mathbb{T}$, and solution set $\mathbb{S}$, respectively. We observe that the model is not converging to the train distribution $\mathbb{T}$, but the solution set $\mathbb{S}$ since $C_\mathrm{KL}^\mathrm{sol}$ is lower than $C_\mathrm{KL}^\mathrm{train}$ and the model is not overfitting.

The fidelity $F$ increases with increasing circuit depth. The best performance was achieved for depth $d\texttt{=}16$, with $\alpha \texttt{=} 0.3$. For circuits with $d$$\in$$\{8,16\}$, the fidelity increases from $\alpha\texttt{=}0.1$ to $\alpha\texttt{=}0.3$ and then decreases. For larger training set sizes, the model has fewer unseen samples $G_\mathrm{new}$ to learn, limiting its ability to generalize. On the contrary, for shallow circuits (i.\,e. $d$$\in$$\{2, 4\}$), $F$ decreases monotonically with increasing training set sizes. These fidelity values indicate a lack of generalization, as the model generates more noisy samples than valid ones.

The exploration $E$ does not provide information about generalization, but the model's ability to generate new valid and invalid samples. We observe that it tends to decrease when increasing the training set size, indicating that fewer generated valid samples are unseen. When the training set is large, there are fewer new samples to generate and, thus, less potential for exploration. 

The normalized coverage $\tilde{C}$ increases with the training set size, meaning the unseen valid samples get more covered. Increasing the training set size enables the model to reach and learn all the valid unique bitstrings.

Lastly, the normalized rate $\tilde{R}$ and precision $P$ exhibit similar behavior as they increase with $\alpha$, indicating the model's ability to generate valid samples not present in the training set. $\tilde{R}$ indicates the proportion of new valid samples, while $P$ accounts for both new valid samples and those present in the training set. For $d\texttt{=}16$, when transitioning from $\alpha\texttt{=}0.1$ to $\alpha\texttt{=}0.3$, the precision $P$ increases from $0.43$$\pm$$0.06$ to $0.64$$\pm$$0.06$, after which it plateaus. The similarity between $\tilde{R}$ and $P$ suggests that the model generates more new valid samples than replicates from the training dataset.
\section{Conclusions and Outlook}
\label{sec:conclusions}
This paper extends the QUARK framework from optimization tasks to a broad class of applications, particularly QML. We demonstrate this improved architecture by implementing a QML benchmarking workflow for generative modeling. In particular, we show how QUARK can support experiments to study the scaling of quantum simulation to multi-GPU environments, assess real quantum hardware and investigate the generalization of QML algorithms. These wide-ranging experiments illustrate the diversity of investigations and metrics in quantum computing research, amplifying the need for a framework like QUARK.

\emph{Future Work:} %
We will continue to extend QUARK and deploy it in further large-scale experimental studies, e.\,g., we will investigate the challenges of learning high-dimensional datasets. Questions arise about the amount of data needed and what training methods perform best on what system configuration. For the shorter term, there is a promising path of exploration of quantum-inspired generative techniques, which include tensor network approaches. We will add tensor network methods, for example, to enhance optimization problems~\cite{https://doi.org/10.48550/arxiv.2101.06250}, to QUARK, making it available to a broad set of existing benchmark kernels. Further, we plan to use QUARK to study the generalization of QML generative algorithms beyond the simple datasets shown in this work.

\section{Acknowledgments}

The authors thank their collaborators in the industry consortium QUTAC for their joint work defining the underlying circuits used in the benchmark runs. The authors additionally thank Marvin Erdmann and  Jernej Rudi Fin\v{z}gar for helpful discussions and manuscript review. CAR and JK are partly funded by the German Ministry for Education and Research (BMB+F) in the project QAI2-Q-KIS under Grant 13N15583. AL and PR are partly funded by the Bavarian Ministry for Economy (StMWi) through the BenchQC research project. The authors generated parts of this text with OpenAI's language-generation models. Upon generation, the authors reviewed, edited, and revised the language.
 \printbibliography{}
\end{document}